\documentclass[aoas,MSNbibl,nameyear,dvips]{arximspdf}
\usepackage{graphicx}

\doi{10.1214/11-AOAS522}
\volume{6}
\issue{2}
\pubyear{2012}
\firstpage{652}
\lastpage{668}

\makeatletter
\def\bbeta{\bolds{\beta}}
\def\M{\mathbf{M}}
\newcommand{\eqref}[1]{(\ref{#1})}
\renewcommand{\citep}[1]{\citeauthor{#1} \citeyear{#1}}

\makeatother

\begin{document}
\begin{frontmatter}

\title{Bayesian hierarchical rule modeling for predicting medical conditions\thanksref{T1,T2}}
\runtitle{Bayesian hierarchical rule modeling}

\thankstext{T1}{Supported in part by a Google Ph.D. fellowship in statistics.}

\thankstext{T2}{Supported in part by NSF Grant IIS-10-53407.}

\begin{aug}
\author[A]{\fnms{Tyler H.} \snm{McCormick}\corref{}\ead[label=e1]{tylermc@u.washington.edu}},
\author[B]{\fnms{Cynthia} \snm{Rudin}\ead[label=e2]{rudin@mit.edu}}
\and
\author[C]{\fnms{David} \snm{Madigan}\ead[label=e3]{madigan@stat.columbia.edu}\ead[label=u1,url]{http://www.stat.columbia.edu/\textasciitilde madigan}}
\runauthor{T. H. McCormick, C. Rudin and D. Madigan}
\affiliation{University of Washington, Massachusetts Institute of
Technology and~Columbia~University}
\address[A]{T. H. McCormick\\
Department of Statistics\\
University of Washington\\
Box 354320\\
Seattle, Washington 98105\\
USA\\
\printead{e1}}
\address[B]{C. Rudin\\
MIT Sloan School of Management, E62-576\\
Massachusetts Institute of Technology\\
Cambridge, Massachusetts 02139\\
USA\\
\printead{e2}}
\address[C]{D. Madigan\\
Department of Statistics\\
Columbia University\\
1255 Amsterdam Ave.\\
New York, New York 10027\\
USA\\
\printead{e3}\\
\printead{u1}}
\end{aug}

\received{\smonth{12} \syear{2010}}
\revised{\smonth{10} \syear{2011}}

%
\begin{abstract}
We propose a statistical modeling technique, called the Hierarchical
Association Rule
Model (HARM), that predicts a patient's possible
future \mbox{medical} conditions given the patient's current and past history of
reported conditions. The core of our technique is a~Bayesian
hierarchical model for selecting predictive association rules (such
as ``\textit{condition \textup{1} and condition \textup{2} $\rightarrow$ condition \textup{3}}'') from
a large set of candidate rules. Because this method ``borrows
strength'' using the conditions of many similar patients, it is able to
provide predictions specialized to any given patient, even when
little information about the patient's history of conditions is
available.
\end{abstract}

%

\end{frontmatter}

\section{Introduction}\label{sec:intro}
The emergence of large-scale medical record databases presents
exciting opportunities for data-based personalized medicine.
Prediction lies at the heart of personalized medicine and in this
paper we propose a~statistical model for predicting patient-level
sequences of medical conditions. We draw on new approaches for
predicting the next event within a ``current sequence,'' given a
``sequence database'' of past event sequences [\citeauthor{RudinLeKoMaSSRN11} (\citeyear{RudinLeKoMaSSRN11,RudinEtAlCOLT11})].
Specifically, we propose the Hierarchical Association Rule Model
(HARM) that generates a set of \textit{association rules} such as
\textit{dyspepsia and epigastric pain} $\rightarrow$
\textit{heartburn}, indicating that dyspepsia and epigastric pain are
commonly followed by heartburn. HARM produces a ranked list of these
association rules. Patients and caregivers can use the rules to
guide medical decisions [see~\citet{Hood:2011ct}, e.g.], while
systems can use predictions to allocate resources~[\citet{Vogenberg:2009uz}]. Built-in explanations represent a particular
advantage of the association rule framework---the rule predicts heartburn
\textit{because} the patient has had dyspepsia and epigastric pain.

In our setup, we assume that each patient visits a healthcare provider
periodically. At each encounter, the provider records time-stamped
medical conditions experienced since the previous encounter.
In this context, we address several prediction problems such as the following:
\begin{itemize}
\item Given data from a sequence of past encounters, predict the next condition
that a patient will report.
\item Given basic demographic information, predict the first condition that
a~patient will report.
\item Given partial data from an encounter (and possibly prior
encounters), predict the next condition.
\end{itemize}

Though medical databases often contain records from thousands or even
millions of patients, most patients experience only a handful of the
massive set of potential conditions. This patient-level sparsity
presents a challenge for predictive modeling.
Our hierarchical modeling approach attempts to address
this challenge by borrowing strength across patients.

The sequential event prediction problem is new, a supervised learning
problem that has been formalized here and by \citeauthor{RudinLeKoMaSSRN11} (\citeyear{RudinLeKoMaSSRN11,RudinEtAlCOLT11}). Rules are particularly useful in
our context: rules yield very interpretable models, and their
conditional probabilities involve few variables and are thus more
reliable to estimate.

The experiments this paper presents indicate that HARM outperforms
several baseline approaches, including a standard ``maximum confidence,
minimum support threshold'' technique used in association rule mining,
and also a nonhierarchical version of our Bayesian method
[from \citeauthor{RudinLeKoMaSSRN11} (\citeyear{RudinLeKoMaSSRN11,RudinEtAlCOLT11})] that ranks rules using
``adjusted confidence.''

More generally, HARM yields a prediction algorithm for sequential
data that can potentially be used for a wide variety of applications
beyond condition prediction. For instance, the algorithm can be
directly used as a recommender system (e.g., for vendors such
as Netflix, amazon.com or online grocery stores such as Fresh Direct
and Peapod). It can be used to predict the next move in a video game
in order to design a more interesting game, or it can be used to
predict the winners at each round of a tournament (e.g., the winners
of games in a football season). All of these applications possess the
same basic structure as the condition prediction problem: a database
consisting of sequences of events, where each event is associated to
an individual entity (medical patient, customer, football team). As
future events unfold in a new sequence, our goal is to predict the
next event.

In Section~\ref{sec:alg} we provide basic definitions and
present our model. In Section~\ref{sec:data} we evaluate
the predictive performance of HARM, along with several baselines\vadjust{\goodbreak}
through experiments on clinical trial data. Section \ref
{SectionRelated} provides related work, and Section~\ref{sec:concl}
provides a discussion and offers potential extensions.


\section{Method} \label{sec:alg}
This work presents a new approach to association rule mining by
determining the ``interestingness'' of rules using a particular
(hierarchical) Bayesian estimate of the probability of exhibiting
condition $b$, given a set of current conditions, $a$.
%
We will first discuss association rule mining and its connection to
Bayesian shrinkage estimators. Then we will present our hierarchical
method for providing personalized condition predictions.
\subsection{Definitions}

An \textit{association rule} in our context is an implication $a
\rightarrow b$ where the left side is a subset of conditions that the
patient has experienced, and~$b$ is a single condition that the patient
has not yet experienced since the last encounter.
Ultimately, we would like to rank rules in terms of ``interestingness''
or relevance for a particular patient at a given time. Using this
ranking, we make predictions of subsequent conditions. Two common
determining factors of the ``interestingness'' of a rule are the
``confidence'' and
``support'' of the rule [\citet{AgrawalEtAl93}; \citet{Piatetsky-Shapiro1991}].

The confidence of a rule $a\rightarrow b$ for a patient is the
empirical probability:
\begin{eqnarray*}
\mathrm{Conf}(a\rightarrow b) &:=& \frac{\mathrm{Number\ of\ times\
conditions\ } a\ \mathrm{and}\ b\ \mathrm{were\ experienced}}{\mathrm
{Number\ of\ times\ conditions\ } a\ \mathrm{were\ experienced}} \\
&:=& \hat{P}(b|a).
\end{eqnarray*}
The support of set $a$ is as follows:
\begin{eqnarray*}
\mathrm{Support}(a) &:=& \mathrm{Number\ of\ times\ conditions\ } a\ \mathrm
{ were\ experienced} \\
&\propto& \hat{P}(a),
\end{eqnarray*}
where $\hat{P}(a)$ is the empirical proportion of times that conditions
$a$ were experienced.
When a patient has experienced a particular set of conditions only a
few times, a new single observation can dramatically alter the
confidence $\hat{P}(b|a)$ for many rules. This problem occurs commonly
in our clinical trial data, where most patients have reported fewer
than 10 total conditions. The vast majority of rule mining algorithms
address this issue with a minimum support threshold to exclude rare
rules, and the remaining rules are evaluated for interestingness
[reviews of interestingness measures include those of \citet{Tan2002}; \citet{GengHamilton07}]. The definition of interestingness is often
heuristic, and is not necessarily a meaningful estimate of $P(b|a)$.

It is well known that problems arise from using a minimum support
threshold. For instance, consider the collection of rules meeting the
minimum support threshold condition. Within this collection, the
confidence alone\vadjust{\goodbreak} should not be used to rank rules: among rules with
similar confidence, the rules with larger support should be preferred.
More importantly, ``nuggets,'' which are rules with low support but very
high confidence, are often excluded by the threshold. This is
problematic, for instance, when a~condition that occurs rarely is
strongly linked with another rare condition; it is essential not to
exclude the rules characterizing these conditions. In our data, the
distribution of conditions has a long tail, where the vast majority of
events happen rarely: out of 1800 possible conditions, 1400 occur less
than 10 times. These 1400 conditions are precisely the ones in danger
of being excluded by a minimum support threshold.

Our work avoids problems with the minimum support threshold by ranking
rules with a shrinkage estimator of $P(b|a)$. These estimators directly
incorporate the support of the rule. One example of such an estimator
is the ``adjusted confidence'' [\citeauthor{RudinEtAlCOLT11} (\citeyear{RudinLeKoMaSSRN11,RudinEtAlCOLT11})]:
\[
\mathrm{AdjConf}(a\,{\rightarrow}\,b, K):= \frac
{\mathrm{Number\ of\ times\ conditions }\ a\ \mathrm{and\ } b\ \mathrm{
were\
experienced}}
{\mathrm{Number\ of\ times\ conditions}\ a\ \mathrm{were\ experienced}+K}.
\]
The effect of the penalty term $K$ is to pull low-support rules toward
the bottom of the list; any rule achieving a high adjusted confidence
must overcome this pull through either a high enough support or a high
confidence. Using the adjusted confidence avoids the problems discussed
earlier: ``interestingness'' is closely related to the conditional
probability $P(b|a)$, and, among rules with equal confidence, the
higher support rules are preferred, and there is no strict minimum
support threshold.

In this work we extend the adjusted confidence model in an important
respect, in that our method shares information across similar patients
to better estimate the conditional probabilities.
The adjusted confidence is a~particular Bayesian estimate of the confidence. Assuming a Beta prior
distribution for the confidence, the posterior mean is
\[
\tilde{P}(b|a):=\frac{\alpha+ \#(a \& b)}{\alpha+\beta+\#a},
\]
where $\#x$ is the support of condition $x$, and $\alpha$ and $\beta$
denote the parameters of the (conjugate) Beta prior distribution.
Our model allows
the parameters of the Binomial to be chosen differently for each
patient and also for each rule. This means that our model can
determine, for instance, whether a~particular patient is more likely
to repeat a~condition that has occurred only once, and also
whether a~particular condition is more likely to repeat than another.


We note that our approach makes no explicit attempt to infer causal
relationships between conditions. The observed associations may
in fact arise from common prior causes such as other conditions or
drugs. Thus, a rule such as \textit{dyspepsia} $\rightarrow$ \textit{heartburn}
does not necessarily imply that successful treatment of dyspepsia will
change the probability of heartburn. Rather, the goal is to
accurately predict heartburn in order to facilitate\vadjust{\goodbreak} effective medical
management. The article of~\citet{citeulike:8505671} contains a more
complete discussion of this distinction.


\subsection{Hierarchical association rule model (HARM)}
\label{sec:model}
For a patient $i$ and a~given rule, $r$, say, we observe $y_{ir}$
co-occurrences (number of times lhs and rhs were experienced), where
there were a total of $n_{ir}$ encounters that include the lhs
($n_{ir}$ is the support for lhs). We model the number of
co-occurrences as $\operatorname{Binomial}(n_{ir},p_{ir})$ and then model
$p_{ir}$ hierarchically to share information across groups of similar
individuals. Define $\M$ as a $I \times D$ matrix of static observable
characteristics for a total of $I$ individuals and $D$ observable
characteristics, where we assume $D>1$ (otherwise we revert back to a
model with a rule-wise adjustment). Each row of $\M$ corresponds to a
patient and each column to a particular characteristic. We define the
columns of $\M$ to be indicators of particular patient categories
(gender, or age between 30 and 40, e.g.), though they could be
continuous in other applications. Let~$\M_i$ denote the $i$th row of the matrix $\M$.
We model the probability for the
$i$th individual and the $r$th rule $p_{ir}$ as
coming from a Beta distribution with parameters $\pi_{ir}$ and $\tau
_{i}$. We then define $\pi_{ir}$ through the regression model $\pi
_{ir}=\exp(\M_i'\bbeta_r+\gamma_i)$, where $\bbeta_r$ defines a vector
of regression coefficients for rule $r$ and $\gamma_i$ is an
individual-specific random effect. More formally, we propose the
following model:
\begin{eqnarray*}
y_{ir}&\sim&\operatorname{Binomial}(n_{ir},p_{ir}),\\
p_{ir}&\sim&\operatorname{Beta}(\pi_{ir}, \tau_{i}),\\
\pi_{ir}&=&\exp(\M_i'\bbeta_{r}+\gamma_i).
\end{eqnarray*}
Under this model,
\[
E(p_{ir}|y_{ir}, n_{ir})=\frac{y_{ir}+\pi_{ir}}{n_{ir}+\pi_{ir}+\tau_{i}},
\]
which is a more flexible form of adjusted confidence. This expectation
also produces nonzero probabilities for a rule even if $n_{ir}$ is zero
(patient $i$ has never reported the conditions on the left-hand side of
$r$ before). This could allow rules to be ranked more highly even if
$n_{ir}$ is zero. The fixed effect regression component, $\M_{i}'\bbeta
_r$, adjusts $\pi_{ir}$ based on the patient characteristics in the $\M
$ matrix. For example, if the entries of $\M$ represented only gender,
then the regression model with intercept $\beta_{r,0} $ would be $\beta
_{r,0}+\beta_{r,1}\mathbf{1}_{\mathrm{male}}$, where $\mathbf
{1}_{\mathrm{male}}$ is one for male respondents and zero for
females. Being male, therefore, has a multiplicative effect of $e^{\beta
_{r,1}}$ on~$\pi_{ir}$. In this example, the $\M_{i}'\bbeta_r$ value is
the same for all males, encouraging similar individuals to have similar
values of $\pi_{ir}$. For each rule $r$, we will use a common prior on
all coefficients in $\bbeta_r$; this imposes a hierarchical structure,
and has the effect of regularizing coefficients associated with rare
characteristics. 

The $\pi_{ir}$'s allow rare but important ``nuggets'' to be
recommended. Even across multiple patient\vadjust{\goodbreak} encounters, many conditions
occur very infrequently. In some cases these conditions may still be
highly associated with certain other conditions. For instance, compared
to some conditions, migraines are relatively rare. Patients who have
migraines, however, typically also experience nausea. A minimum support
threshold algorithm might easily exclude the rule ``migraines
$\rightarrow$ nausea'' if a patient hasn't experienced many migraines in
the past. This is especially likely for patients who have few
encounters. In our model, the $\pi_{ir}$ term balances the
regularization imposed by~$\tau_{i}$ to, for certain individuals,
increase the ranking of rules with high confidence but low support. The
$\tau_{i}$ term reduces the probability associated with rules that have
appeared few times in the data (low support), with the same effect as
the penalty term $(K)$ in the adjusted confidence. Unlike the
cross-validation or heuristic strategies suggested in \citeauthor{RudinLeKoMaSSRN11}
(\citeyear{RudinLeKoMaSSRN11,RudinEtAlCOLT11}), we estimate $\tau_{i}$ as part of
an underlying statistical model. Within a given rule, we assume $\tau
_{i}$ for every individual comes from the same distribution. This
imposes additional structure across individuals, increasing stability
for individuals with few observations.

It remains now to describe the precise prior structure on the
regression parameters and hyperparameters. We assign Gaussian priors
with mean $0$ and variance $\sigma^2_{\tau}$ to the $\tau$ on the log
scale. Since any given patient is unlikely to experience a specific
medical condition, the majority of probabilities are close to zero.
Giving $\tau_{i}$ a prior with mean zero improves stability by
discouraging excessive penalties.
We assign all elements $\beta_{r,d}$ of vectors~$\bbeta_r$ a~common
Gaussian prior on the log scale with mean $\mu_{\bbeta}$ and variance
$\sigma^2_{\bbeta}$. We also assume each $\gamma_i$ comes from a
Gaussian distribution on the log scale with common mean $\mu_\gamma$
and variance $\sigma^2_\gamma$. 
Each individual has their own $\gamma_i$ term, which permits
flexibility among individuals; however, all of the $\gamma_i$ terms
come from the same distribution, which induces dependence between
individuals. We assume diffuse uniform priors on the hyperparameters
$\sigma^2_{\tau}$,~$\mu_{\bbeta}$ and~$\sigma^2_{\bbeta}$. Denote
$\mathbf{Y}$ as the matrix of $y_{ir}$ values, $\mathbf{N}$ as the
matrix of $n_{ir}$ values, and $\bbeta$ as the collection of $\bbeta
_1,\ldots, \bbeta_R$. The prior assumptions yield the following posterior:
\begin{eqnarray*}
p,\pi, \tau, \bbeta|\mathbf{Y}, \mathbf{N}, \M&\propto& \prod_{i=1}^I
\prod_{r=1}^R
p_{ir}^{y_{ir}+\pi_{ir}}(1-p_{ir})^{n_{ir}-y_{ir}+\tau_{i}}\\
&&{}\times\prod_{r=1}^R\prod_{d=1}^D\operatorname{Normal}(\log(\beta_{r,d})|\mu
_{\bbeta},\sigma^2_{\bbeta})\\
&&{}\times\prod_{i=1}^I\operatorname{Normal}(\log({\gamma}_i)|\mu_{\gamma},\sigma
^2_{\gamma})\operatorname{Normal}(\log(\tau_{i})|0,\sigma^2_{\tau}).
\end{eqnarray*}

\begin{figure}

\includegraphics{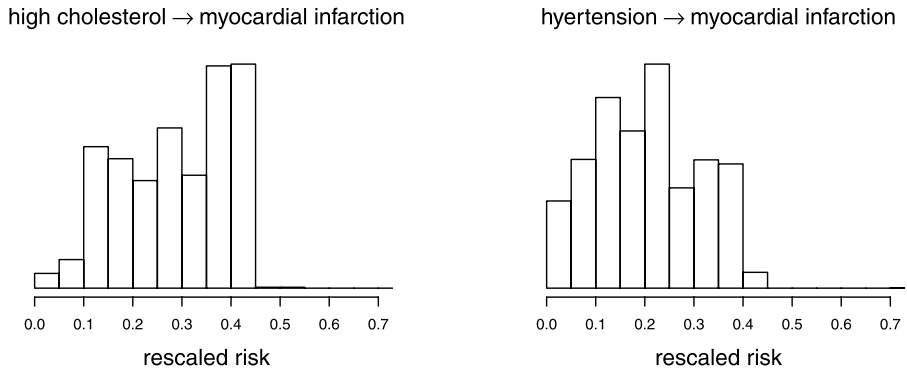}

\caption{Approximate posterior of rescaled risk for two rules. These
are histograms of the posterior means for the set of patients.}
\label{fig:posthist}
\end{figure}

HARM produces draws from the (approximate) posterior distribution for
each~$p_{ir}$. Since these terms will be used for ranking the rules, we
refer to them as rescaled risk.\vadjust{\goodbreak} We also note that, even though the
$p_{ir}$'s represent probabilities in our model, they are not
interpretable as the probability that a patient will have a given
condition at the next visit to a provider (since our model is not
calibrated to time between visits). Figure~\ref{fig:posthist} shows
estimates of the posterior rescaled risk for \textit{high cholesterol}
$\rightarrow$ \textit{myocardial infarction} and \textit{hypertension}
$\rightarrow$ \textit{myocardial infarction}. Comparing the distributions
of related rules can often provide insights into associations in the
data, as we demonstrate in Section~\ref{subsection:knownassoc}.%
In the context of medical condition prediction, these rescaled risks
are of interest and we analyze our estimates of their full posterior
distributions in Section \ref{subsection:knownassoc}. To rank
association rules for the purpose of prediction, however, we need a
single estimate for each probability (rather than a full distribution),
which we chose as the posterior mean. In practice, we suggest
evaluating the mean as well other estimators for each rescaled risk
(the mode or median, e.g.) and selecting the one with the best
performance in each particular application. We carry out our
computations using a Gibbs sampling algorithm, provided in Figure \ref
{FigureAlg}.

%
\begin{figure}
\hspace*{-160pt}For a suitably initialized chain, at step $v$:
\begin{enumerate}[10.]
\item[1.] Update $p_{ir}$ from the conjugate Beta distribution given $\pi
_{ir}, \tau_{i}, \mathbf{Y}, \mathbf{N}, \M.$
\item[2.] Update $\tau_{i}$ using a Metropolis step with proposal $\tau
_i^*$ where
\[
\log(\tau_{i}^*)\sim\mathrm{N}\bigl(\tau_{i}^{(v-1)},(\mbox{scale of jumping dist})\bigr).
\]
\item[3.] For each rule, update the vector $\bolds{\beta}_{r}$ using a
Metropolis step with
\[
\log(\bolds{\beta}_r^*)\sim\mathrm{N}\bigl(\bolds{\beta
}_r^{(v-1)},(\mbox{scale of jumping dist})\bigr).
\]
\item[4.] Update $\gamma_{i}$ using a Metropolis step with
\[
\log(\gamma_i^*)\sim\mathrm{N}\bigl(\gamma_i^{(v-1)},(\mbox{scale of jumping dist})\bigr).
\]
\item[5.] Update $\pi_{ir}=\exp(\M_i'\bbeta_{r}+\gamma_i)$.
\item[6.] Update $\mu_{\bbeta}\sim\mathrm{N}(\hat{\mu}_{\bbeta},\sigma_{\beta
}^2)$ where
\[
\hat{\mu}_{\bbeta}=\biggl(\frac{1}{D+R}\biggr)\sum_{r=1}^R\sum_{d=1}^D
{\beta_{r,d}}.
\]
\item[7.] Update $\sigma_{\bbeta}^2 \sim\operatorname{Inv-}\chi^2(d-1,\hat{\sigma
}^2_{\bbeta})$ where
\[
\hat{\sigma}^2_{\bbeta}=\biggl(\frac{1}{D+R-1}\biggr)\sum_{r=1}^R\sum
_{d=1}^D(\beta_{r,d}-\mu_{\bbeta})^2.
\]
\item[8.] Update $\sigma_{\tau}^2 \sim\operatorname{Inv-}\chi^2(I-1,\hat{\sigma
}^2_{\tau})$ where $\hat{\sigma}^2_{{\tau}}=\frac{1}{I-1}\sum
_{i=1}^I(\tau_i-\mu_{\tau})^2$.
\item[9.] Update $\mu_{\gamma}\sim\mathrm{N}(\hat{\mu}_{\gamma},\sigma_{\gamma
}^2)$ where $\hat{\mu}_{\gamma}=\frac{1}{I}\sum_{i=1}^I {\gamma_i}$.
\item[10.] Update $\sigma_{\gamma}^2 \sim\operatorname{Inv-}\chi^2(I-1,\hat{\sigma
}^2_{\gamma})$ where $\hat{\sigma}^2_{\gamma}=\frac{1}{I-1}\sum
_{i=1}^I(\gamma_i-\mu_{\gamma})^2$.
\end{enumerate}
\caption{Gibbs sampling algorithm for the hierarchical Bayesian
association rule modeling for sequential event prediction (HARM).}\label
{FigureAlg}
\end{figure}

\subsection{Approximate updating} \label{onlineupdating}
Given a batch of data, HARM makes predictions based on the posterior
distributions of the $p_{ir}$'s. Since the posteriors are not available
in closed form, we need to iterate the algorithm in Figure~\ref
{FigureAlg} to convergence in order to make predictions. Each time the
patient visits the physician, each $p_{ir}$ could be updated by again
iterating the algorithm in Figure~\ref{FigureAlg} to convergence.
In some applications, new data continue to arrive frequently, making it
impractical to compute approximate posterior distributions using the
algorithm in Figure~\ref{FigureAlg} for each new encounter. In this
section we provide an approximate updating scheme to incorporate new
patient data after an initial batch of encounters has already been
processed. The approximate scheme can be used for real-time online updating.

Beginning with an initial batch of data, we run the algorithm in
Figure~\ref{FigureAlg} to convergence in order to obtain\vadjust{\goodbreak} $\hat{\tau
}_{i}$ and $\hat{\pi}_{ir}$, which are defined to be the posterior
means of the estimated distributions for $\tau_{i}$ and $\pi_{ir}$. The
approximate updating scheme keeps $\tau_i$ and $\pi_{ir}$ fixed to be
$\hat{\tau}_{i}$ and $\hat{\pi}_{ir}$.
Given that up to encounter $e-1$ we have observed $y^{(e-1)}_{ir}$ and
$n^{(e-1)}_{ir}$, we are presented with new observations that have
counts $y^{(\mathrm{newobs.})}_{ir}$ and $n^{(\mathrm{newobs.})}_{ir}$
so that $y^{(e)}_{ir} = y^{(e-1)}_{ir} + y^{(\mathrm{newobs.})}_{ir}$
and $n^{(e)}_{ir} = n^{(e-1)}_{ir} + n^{(\mathrm{newobs.})}_{ir}$. In
order to update the probability estimates to reflect our total current
data, $y^{(e)}_{ir}$, $n^{(e)}_{ir}$, we will use the following relationship:
\begin{eqnarray*}\label{eq:profile}
P\bigl(p_{ir}|y^{(e)}_{ir}, n^{(e)}_{ir},\hat{\tau}_{i},\hat{\pi
}_{ir}\bigr)&\propto&P\bigl(y^{(\mathrm{newobs.})}_{ir}| n^{(\mathrm
{newobs.})}_{ir}, p_{ir}\bigr)\\
&&{}\times P\bigl(p_{ir}|y^{(e-1)}_{ir}, n^{(e-1)}_{ir},\hat{\tau}_{i},\hat{\pi}_{ir}\bigr).
\end{eqnarray*}
%
The expression $P(p_{ir}|y^{(e-1)}_{ir}, n^{(e-1)}_{ir}, \hat{\tau
}_{i},\hat{\pi}_{ir})$ is the posterior up to encounter $e-1$ and has a
Beta distribution. The likelihood of the new observations,
$P(y^{(\mathrm{newobs.})}_{ir}| n^{(\mathrm{newobs.})}_{ir}, p_{ir})$,
is Binomial. Conjugacy implies that the updated posterior also has a
Beta distribution. In order to update the probability estimates for our
hierarchical model, we use the expectation of this distribution, that is,
\[
E\bigl(p_{ir}|y_{ir}^{(e)}, n_{ir}^{(e)},\hat{\tau}_{i},\hat{\pi}_{ir}\bigr)=\frac
{y^{(e-1)}_{ir}+y^{\mathrm{newobs.}}_{ir}+\hat{\pi
}_{ir}}{n^{(e-1)}_{ir}+n^{\mathrm{newobs.}}_{ir}+\hat{\pi}_{ir}+\hat
{\tau}_{i}}.
\]

%

\section{Application to repeated patient encounters}
\label{sec:data}
We present results of\break HARM, with the approximate updating scheme in
Section \ref{onlineupdating}, on co-prescrib\-ing data from a large
clinical trial. In the trial, each patient visits a healthcare provider
periodically. At each encounter, the provider records time-stamped
medical conditions (represented by MedDRA terms) experienced since the
previous encounter. Thus, each encounter is associated with a sequence
of medical conditions.
These data are from around 42,000 patient encounters from about 2,300
patients, all at least 40 years old.
The matrix of observable characteristics encodes the basic demographic
information: gender, age, and ethnicity. For each patient we have a
record of each medication prescribed and the condition/chief complaint
(back pain, asthma, etc.) that warranted the prescription. We chose to
predict patient complaints rather than prescriptions since there are
often multiple prescribing options (medications) for the same complaint.
Some patients had preexisting conditions that continued throughout the
trial. For these patients, we include these preexisting conditions in
the patient's list of conditions at each encounter.
Other patients have recurrent conditions for which we would like to
predict the occurrences.
If a patient reports the same condition more than once during the same
thirty day period, we only consider the first occurrence of the
condition at the first report. If the patient reports the condition
once and then again more than thirty days later, we consider this two
separate incidents.\looseness=-1

As covariates, we used age, gender, race and drug/placebo (an
indicator of whether the patient was in the treatment or control group
for the clinical trial). We fit age using a series of indicator
variables corresponding to four groups (40--49, 50--59, 60--69, 70$+$). We
included all available covariates in our simulation studies. In
practice, model selection will likely be essential to select the best
subset of covariates for predictive performance. We discuss covariate
selection in further detail in the supplemental
article~[\citet{onlinesup}].

Our experiments consider only the marginal probabilities (support) and
probabilities conditional on one previous condition. Thus, the
left-hand side of each rule contains either 0 items or 1 item. In our
simulations, we used chains of 5000 iterations, keeping every 10th
iteration to compute the mean we used for ranking and discarding the
first thousand iterations.\vadjust{\goodbreak}

In Section \ref{subsection:prediction} we present experimental results
to compare the predictive performance of our model to other rule mining
algorithms for this type of problem. In Section \ref
{subsection:knownassoc} we use the probability estimates from the model
to demonstrate its ability to find new associations; in particular, we
find associations that are present in the medical literature but that
may not be obvious by considering only the raw data.

\subsection{Predictive performance}\label{subsection:prediction}
We selected a sample of patients by assigning each patient a random
draw from a Bernoulli distribution with success probability selected to
give a sample of patients on average around 200. For each patient we
drew uniformly an integer $t_{i}$ between 0 and the number of
encounters for that patient. We ordered the encounters chronologically
and used encounters $1$ through~$t_{i}$ as our training set and the
remaining encounters as the test set. Through this approach, the
training set encompasses the complete set of encounters for some
patients (``fully observed''), includes no encounters for others (``new
patients''), and a partial encounter history of the majority of the test
patients (``partially-observed patients''). We believe this to be a
reasonable approximation of the context where this type of method would
be applied, with some patients having already been observed several
times and other new patients entering the system for the first time. We
evaluated HARM's predictive performance using a combination of common
and rare conditions. For each run of the simulation, we use the 25 most
popular conditions, then randomly select an additional 25 conditions
for a~total of 50.

The algorithm was used to iteratively predict the conditions revealed
at each encounter. For each selected patient, starting with the first
test encounter, and prior to that encounter's first condition being
revealed, the algorithm made a prediction of $c$ possible conditions,
where $c=3$. Note that to predict the very first condition for a given
patient when there are no previous encounters, the recommendations come
from posterior means of the coefficients estimated from the training
set. The algorithm earned one point if it recommended the current
condition before it was revealed, and no points otherwise. Then,
$y_{ir}$ and $n_{ir}$ were updated to include the revealed condition.
This process was repeated for the patient's remaining conditions in the
first encounter, and repeated for each condition within each subsequent
encounter. We then moved to the next patient and repeated the procedure.

The total score of the algorithm for a given patient was computed as
the total number of points earned for that patient divided by the total
number of conditions experienced by the patient. The total score of the
algorithm is the average of the scores for the individual patients.
Thus, the total score is the average proportion of correct predictions
per patient. We repeated this entire process (beginning with selecting
patients) 500 times and recorded the distribution over the 500 scores.
We compared the performance of HARM\vadjust{\goodbreak} (using the same scoring system)
against an algorithm that ranks rules by adjusted confidence, for
several values of $K$. We also compared with the ``max confidence
minimum support threshold'' algorithm for different values of the
support threshold $\theta$, where rules with support below $\theta$ are
excluded and the remaining rules are ranked by confidence. For both of
these algorithms, no information across patients is able to be used.

\begin{figure}

\includegraphics{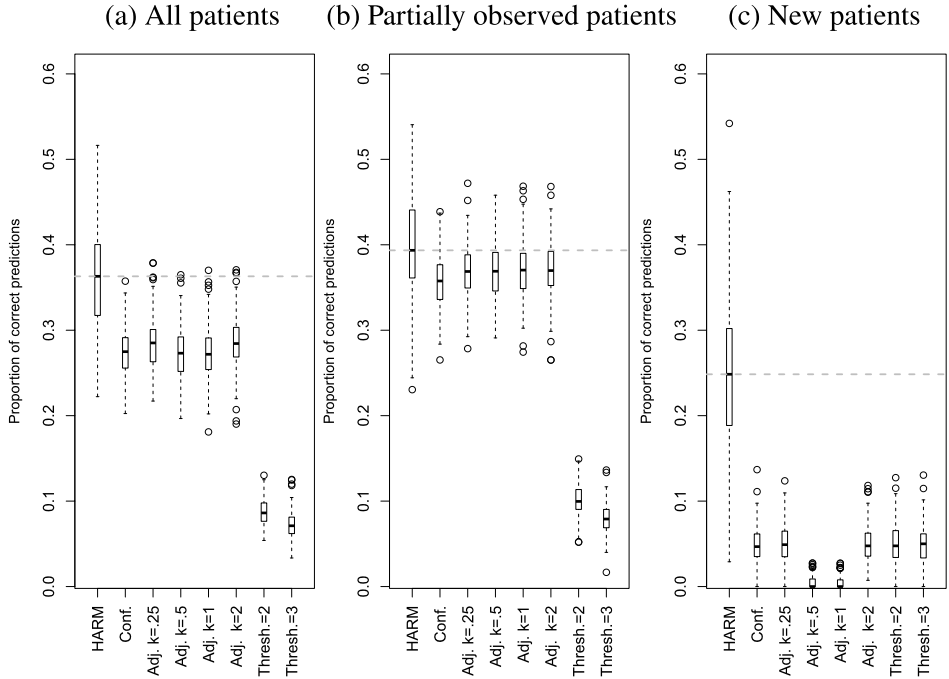}

\caption{Predictive performance for (\textup{a}) all patients, (\textup{b})
partially-observed patients, (\textup{c})~new patients. Each boxplot represents
the distribution of scores over 500 runs. For~(\textup{a}), each run's score (an
individual point on a boxplot) is based on a sample of approximately
200 patients. For (\textup{b}) and (\textup{c}), each point is based on a subset of these
$\sim$200 patients.}
\label{fig:full}
\end{figure}

Figure~\ref{fig:full} 
shows the results, as boxplots of the
distribution of scores for the entire collection of partially-observed,
fully observed and new patients. 
Paired $t$-tests comparing the mean proportion of correct predictions
from HARM to each of the alternatives had $p$-values for a significant
difference in our favor less than $10^{-15}$. In other words, HARM has
statistically superior performance over all $K$ and $\theta$,
{that is}, better performance than either of the two algorithms even if
their parameters $K$ and $\theta$ had been tuned to the best possible value.
For all four values of $K$ for the adjusted confidence, performance was
slightly better than for the plain confidence ($K=0$). The ``max
confidence minimum support threshold'' algorithm (which is a standard
approach to association rule mining problems) performed poorly for
minimum support thresholds of 2 and 3. This poor performance is likely
due to the sparse information we have for each patient. Setting a
minimum support threshold as low as even two eliminates many potential
candidate rules from consideration.


%


The main advantage of our model is that it shares information across
patients in the training set. This means that in early stages where the
observed~$y_{ir}$ and $n_{ir}$ are small, it may still be possible to
obtain reasonably accurate probability estimates,
since when patients are new, our recommendations depend heavily on the
behavior of previously observed similar patients.
This advantage is shown explicitly through Figures \ref{fig:full}(b)
and \ref{fig:full}(c), which pertain to partially-observed and new
patients, respectively.
The advantage of HARM over the other methods is more pronounced for new
patients: in cases where there are no data for each patient, there is a
large advantage to sharing information.
We performed additional simulations which further illustrate this point
and are presented in the supplement [\citet{onlinesup}].

\subsection{Association mining}
\label{subsection:knownassoc}
The conditional probability estimates from our model are also a way of
mining a large and highly dependent set of associations.

%
\begin{figure}
\centering
\begin{tabular}{@{}cc@{}}

\includegraphics{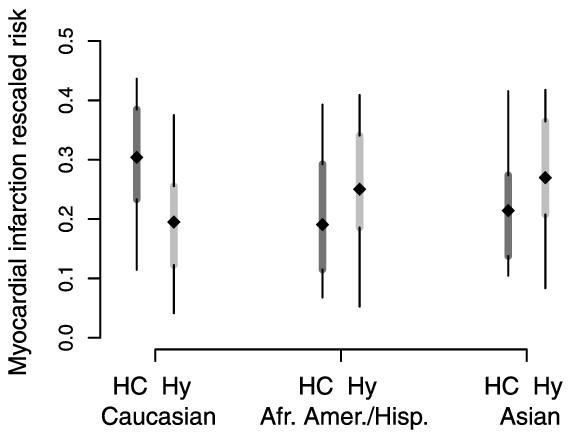}
 & \includegraphics{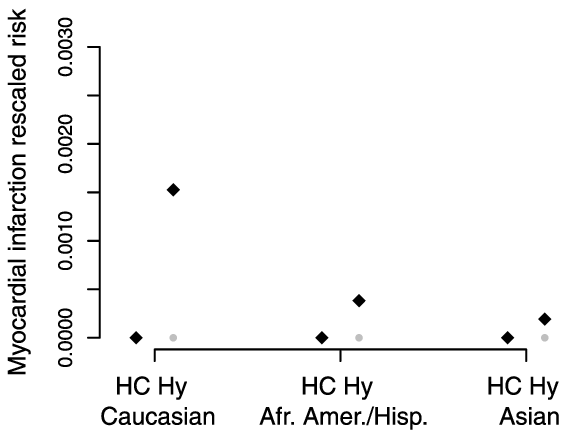}\\
\footnotesize{(a)} & \footnotesize{(b)}
\end{tabular}
\caption{Propensity of myocardial infarction in patients who have
reported high cholesterol or hypertension using (\textup{a}) HARM and (\textup{b})
(unadjusted) confidence. For each demographic group, high cholesterol
(HC) is on the left and hypertension (Hy) is on the right. Thick lines
represent the middle half of the posterior mean propensities for
respondents in the indicated demographic group. Outer lines represent
the middle 90$\%$ and dots represent the mean.
The vast majority of patients did not experience a myocardial
infarction, which places the middle 90$\%$ of the distributions in plot
(\textup{b}) approximately at zero.}
\label{fig:race}
\end{figure}

\textit{Ethnicity, high cholesterol or hypertension
$\rightarrow$ myocardial infarction}:
Figure~\ref{fig:race}(a) shows the distribution of posterior mean
propensity for myocardial infarction (heart attack) given two
conditions previously reported as risk factors for myocardial
infarction: high cholesterol and hypertension [see \citet{cdc} for a
recent review].
Each bar in the figure corresponds to the set of 
respondents in a specified ethnic group. For Caucasians, we typically
estimate a higher probability of myocardial infarction in patients who
have previously had high cholesterol. In African Americans/Hispanics
and Asian patients, however, we estimate a generally higher probability
for patients who have reported hypertension.
This distinction demonstrates the flexibility of our method in
combining information across respondents who are observably similar.
Race-ethnic differences in risk
factors for coronary heart disease have attracted considerable
attention in the medical literature [see, e.g.,~\citet
{Rosamond06022007} or~\citet{Willey:2011jq}].

As a comparison, we also included the same plot using (unadjusted)
confidence, in Figure \ref{fig:race}(b). In both Figure \ref
{fig:race}(a) and Figure \ref{fig:race}(b), the black dots are the mean
across all the patients, which
are not uniformly at zero because there were some cases of myocardial
infarction and hypertension or high cholesterol.
In Figure \ref{fig:race}(b), the colored, smaller dots represent the
rest of the middle 90$\%$ of the distribution, which appears to be at
zero in plot (b) since the vast majority of patients did not have a
myocardial infarction at all, so even fewer had a myocardial infarction
after reporting hypertension or high cholesterol.

\begin{figure}
\centering
\begin{tabular}{@{}cc@{}}

\includegraphics{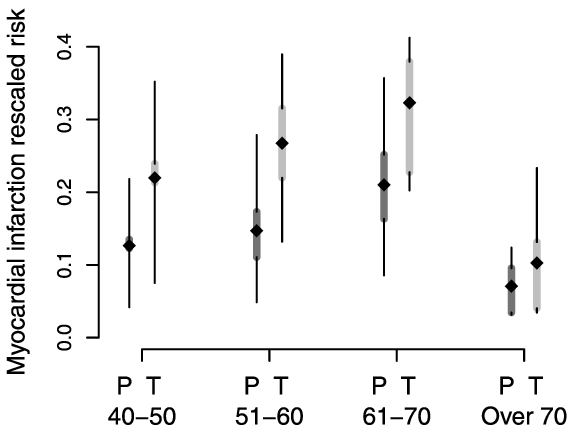}
 & \includegraphics{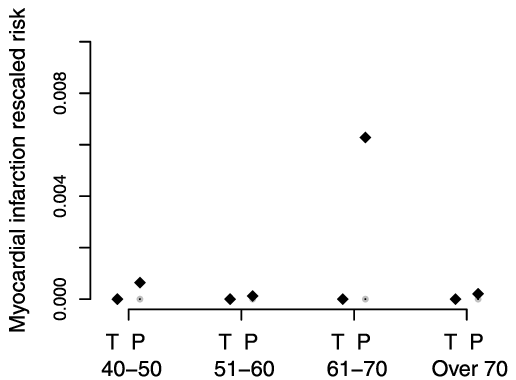}\\
\footnotesize{(a)} & \footnotesize{(b)}
\end{tabular}
\caption{Propensity of myocardial infarction in patients who have
reported hypertension, estimated by (\textup{a}) HARM and (\textup{b}) (unadjusted)
confidence. For each demographic group, the placebo (P) is on the left
and the treatment medication (T) is on the right. Thick lines represent
the middle half of the posterior mean propensities for respondents in
the indicated demographic group. Outer lines represent the middle 90$\%
$ and dots represent the mean.
Overall the propensity is higher for individuals who take the study
medication than those who do not.}
\label{fig:trt}
\end{figure}

\begin{figure}[b]
\centering
\begin{tabular}{@{}cc@{}}

\includegraphics{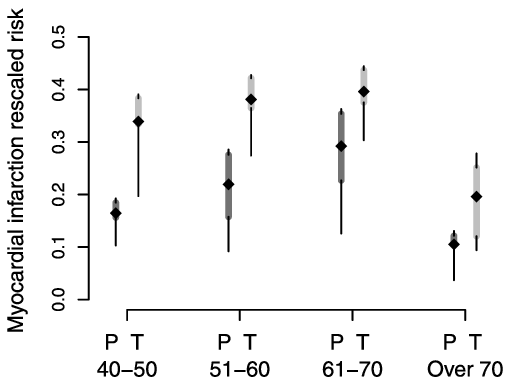}
 & \includegraphics{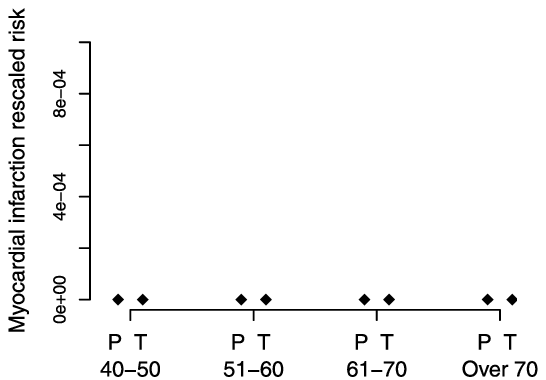}\\
\footnotesize{(a)} & \footnotesize{(b)}
\end{tabular}
\caption{Propensity of myocardial infarction in patients who have
reported high cholesterol, estimated by (\textup{a}) HARM and (\textup{b}) (unadjusted)
confidence.
}
\label{fig:trt2}
\end{figure}

\textit{Age, high cholesterol or hypertension, treatment
or placebo $\rightarrow$ myocardial infarction}:
Since our data come from a clinical trial, we also included an
indicator of treatment vs. placebo condition in the hierarchical
regression component of HARM.
Figures~\ref{fig:trt} and~\ref{fig:trt2} display the posterior means of
propensity of myocardial infarction for respondents separated by age
and treatment condition. Figure \ref{fig:trt} considers patients who
have reported hypertension, Figure \ref{fig:trt2} considers patients
who have reported high cholesterol. In both Figure \ref{fig:trt} and
Figure \ref{fig:trt2}, it appears that the propensity of myocardial
infarction predicted by HARM is greatest for individuals between 50 and
70, with the association again being stronger for high cholesterol than
hypertension.\looseness=-1

%
%
%

For both individuals with either high cholesterol or hypertension, use
of the treatment medication was associated with increased propensity of
myocardial infarction. This effect is present across nearly every age
category. The distinction is perhaps most clear among patients in their
fifties in both Figure \ref{fig:trt} and Figure~\ref{fig:trt2}. The
treatment product in this trial has been linked to increased risk of
myocardial infarction in numerous other studies. The product was
eventually withdrawn from the market by the manufacturer because of its
association with myocardial infarctions.

The structure imposed by our hierarchical model gives positive
probabilities even when no data are present in a given category; in
several of the categories, we observed no instances of a myocardial
infarction, 
so estimates using only the data cannot differentiate between the
categories in terms of risk for myocardial infarction, as particularly
illustrated through Figure~\ref{fig:trt2}(b).\looseness=-1

\section{Related works}\label{SectionRelated}
Four relevant works on Bayesian hierarchical modeling and recommender\vadjust{\goodbreak}
systems are those of \citet{dumouchel2001},
\citet{Breese98empiricalanalysis}, \citet
{Condliff99bayesianmixed-effects} and \citet{AgarwalZhMa11}. \citet
{dumouchel2001}
deal with the identification of interesting itemsets (rather than
identification of rules). Specifically, they model the ratio of
observed itemset frequencies to baseline frequencies computed under a
particular model for independence. Neither~\citet
{Condliff99bayesianmixed-effects} nor~\citet{Breese98empiricalanalysis}
aim to model repeat purchases (recurring conditions). \citet
{Breese98empiricalanalysis} use Bayes\-ian
methods to cluster users, and also suggest a~Bayes\-ian network.
\citet{Condliff99bayesianmixed-effects} present a hierarchical
Bayesian approach to collaborative filtering that ``borrows strength''
across users. \citet{AgarwalZhMa11} also build a personalized
recommender system that models item-item similarities.
Their model uses logistic regression for estimating $p_{ir}$ rather
than using $\pi_{ir}$ and $\tau_{i}$. This has the advantage of being a
simpler hierarchical model, but loses the interpretability our model
has through using association rules. It also loses the potential
advantage of estimating only conditional probabilities involving few
variables.

As far as we know, the line of work by~\citet{davis10} is the first to
use an
approach from recommender systems to predict medical conditions, though
in a
completely different way than ours; it is based on vector similarity,
in the
same way as~\citet{Breese98empiricalanalysis}. [Also see references
in~\citet{davis10} for
background on collaborative filtering.] Also in the context of learning
in medical problems,~\citet{Gopalakrishnan:2010tm} used decision rules
chosen using a Bayesian network to predict disease state from biomarker
profiling studies.


\section{Conclusion and future work}
\label{sec:concl}
We have presented a hierarchical model for ranking association rules
for sequential event prediction. The sequential nature of the data is
captured through rules that are sensitive to time order, that is,
$a\rightarrow b$ indicates conditions $a$ are followed by conditions
$b$. HARM uses information from observably similar individuals to
augment the (often sparse) data on a particular individual; this is how
HARM is able to estimate probabilities $P(b|a)$ before conditions $a$
have ever been reported. In the absence of data, hierarchical modeling
provides structure. As more data become available, the influence of the
modeling choices fade as greater weight is placed on the data. The
sequential prediction approach is especially well suited to medical
condition prediction, where experiencing two conditions in succession
may have different clinical implications than experiencing either
condition in isolation.

Model selection is important for using our method in practice. There
are two types of model selection required for HARM: the choice of
covariates encoded by the matrix $\M$, and the collection of available
rules. For the choice of covariates in $\M$, standard feature selection
methods can be used, for instance, a forward stagewise procedure where
one covariate at a~time is added as performance improves, or a~backward
stagewise method where features are iteratively removed. Another
possibility is to combine covariates, potentially through a method
similar to model-based clustering~[\citet{MClust}]. To perform model
selection on the choice of rules, it is possible to construct analogous
``rule selection'' methods as one might use for a set of covariates. A
forward stagewise procedure could be constructed, where the set of
rules is gradually expanded as prediction performance increases.
Further, it is possible to combine a set of rules into a single rule as
in model-based clustering; {for example}, rather than separate
rules where the left side is either ``dorsal pain,'' ``back pain,'' ``low
back pain,'' or ``neck pain,'' we could use simply ``back or neck pain''
for all of them.\vadjust{\goodbreak}

Another direction for future work is to incorporate a model for
higher-order dependence, along the line of work by \citet{MTD02}. An
algorithm for sequential event prediction is presented in ongoing work
[\citet{LethamEtAl11}], which is loosely inspired by the ideas of \citet
{MTD02}, but does not depend on association rules. A third potential
future direction is to design a more sophisticated online updating
procedure than the one in Section~\ref{onlineupdating}. It may be
possible to design a procedure that approximately updates all of the
hyperparameters as more data arrive.

\section*{Acknowledgments}
We would like to thank the
Editor, Associate Editor and referee for helpful comments and suggestions.

\begin{supplement}[id=suppA]
\stitle{Additional simulation results}
\slink[doi]{10.1214/11-AOAS522SUPP} 
\slink[url]{http://lib.stat.cmu.edu/aoas/522/supplement.pdf}
\sdatatype{.pdf}
\sdescription{In the supplement we present additional simulation
results which speak to the performance of HARM.}
\end{supplement}

%


\printaddresses

\end{document}